\documentclass[]{rnaastex}

\usepackage{graphicx}
\usepackage[suffix=]{epstopdf}
\usepackage{natbib}
\usepackage{amsmath}
\usepackage{xspace}

\newcommand{\Kepler}{\textsl{Kepler}\xspace}

\begin{document}

\title{Infrared Flares from M Dwarfs: a Hinderance to\\ Future Transiting Exoplanet Studies}

\correspondingauthor{James. R. A. Davenport}
\email{jrad@uw.edu}

\author{James. R. A. Davenport}
\altaffiliation{NSF Astronomy and Astrophysics Postdoctoral Fellow}
\altaffiliation{DIRAC Fellow}
\affiliation{Department of Physics \& Astronomy, Western Washington University, 516 High St., Bellingham, WA 98225, USA}
\affiliation{Department of Astronomy, University of Washington, Seattle, WA 98195, USA}


\section{} 

Many current and future exoplanet missions are pushing to infrared (IR) wavelengths where the flux contrast between the planet and star is more favorable \citep{deming2009}, and the impact of stellar magnetic activity is decreased. Indeed, a recent analysis of starspots and faculae found these forms of stellar activity do not substantially impact the transit signatures or science potential for FGKM stars with JWST \citep{zellem2017}. However, this is not true in the case of flares, which I demonstrate can be a hinderance to transit studies in this note. 

\citet{tofflemire2012} noted the potential impact of flares for transit studies in the IR, and found upper-limits for IR flare flux for mid-M stars. \citet{davenport2012} created peak flare flux conversions between optical and IR bands for M0--M6 stars, and noted the IR flare flux contrast is strongest for later spectral types. Flares have already been a roadblock for detecting transits for stars in the optical for M dwarfs \citep[e.g. Proxima b:][]{davenport2016a, kipping2017}. Late M dwarfs, while appealing targets for transit discoveries and transit-transmission-spectroscopy studies, are therefore the most susceptible to contamination by flares in the IR.

\begin{figure*}[h!]
\begin{center}
\includegraphics[height=2in]{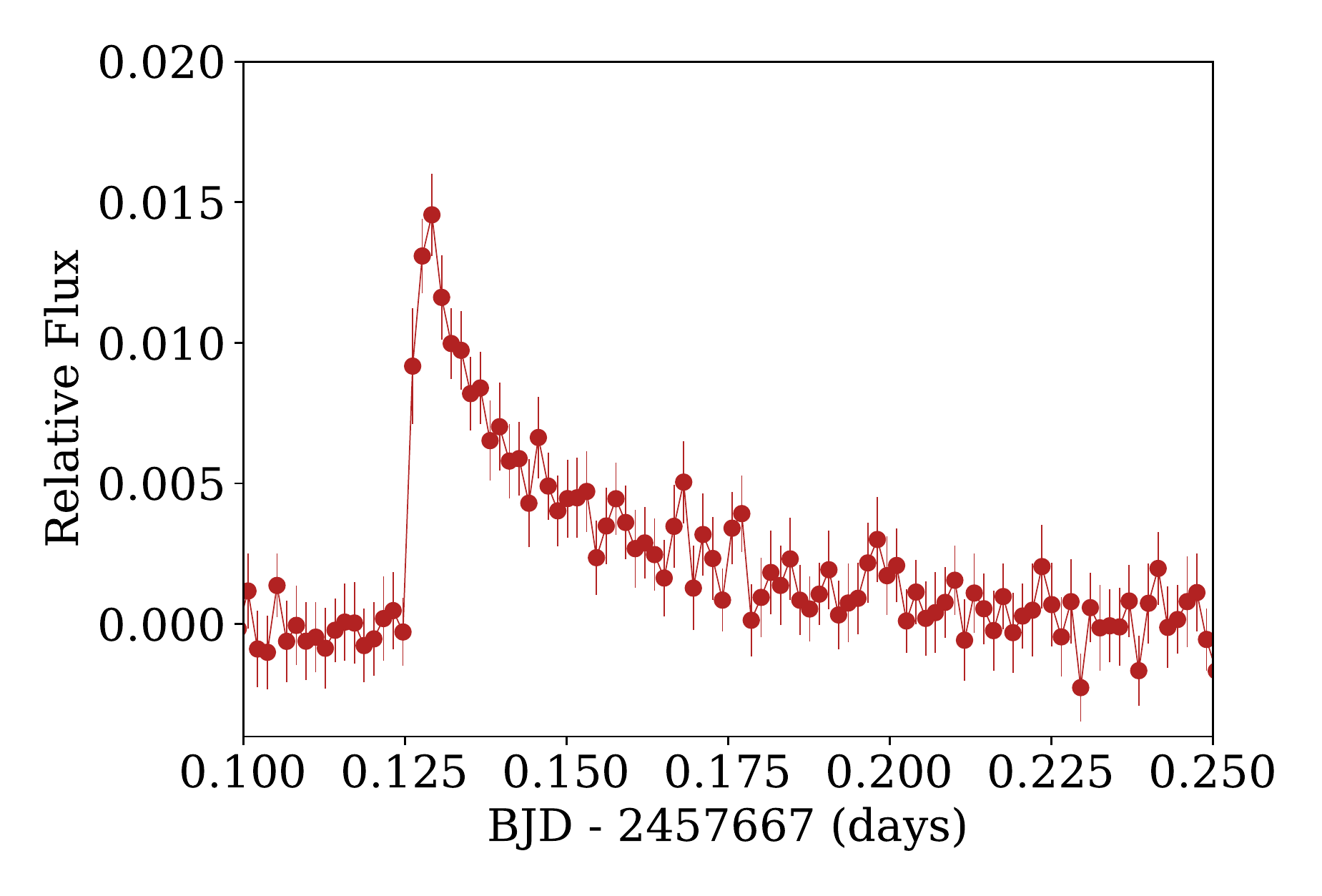}
\includegraphics[height=2in]{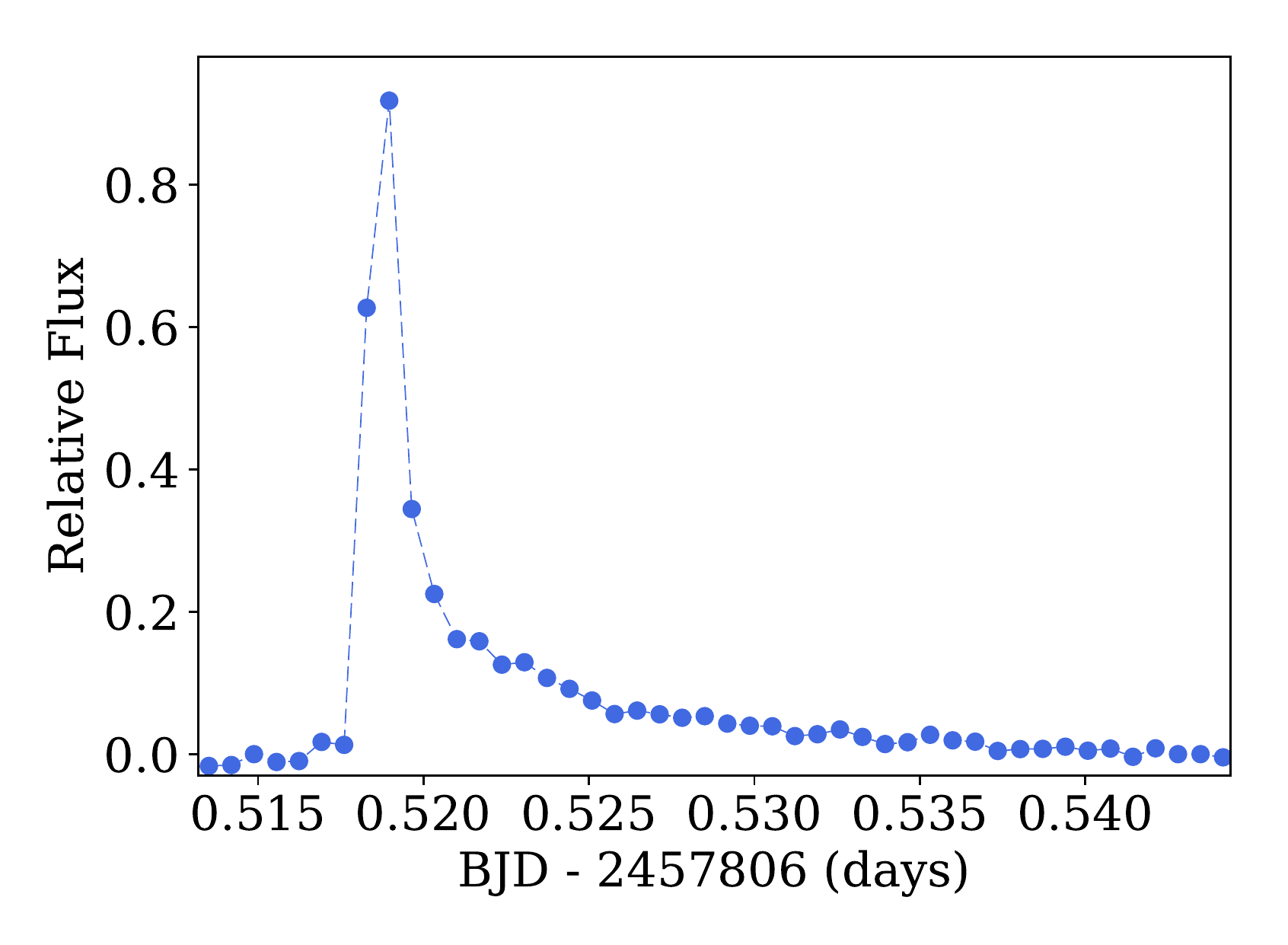}\\
\includegraphics[width=4in]{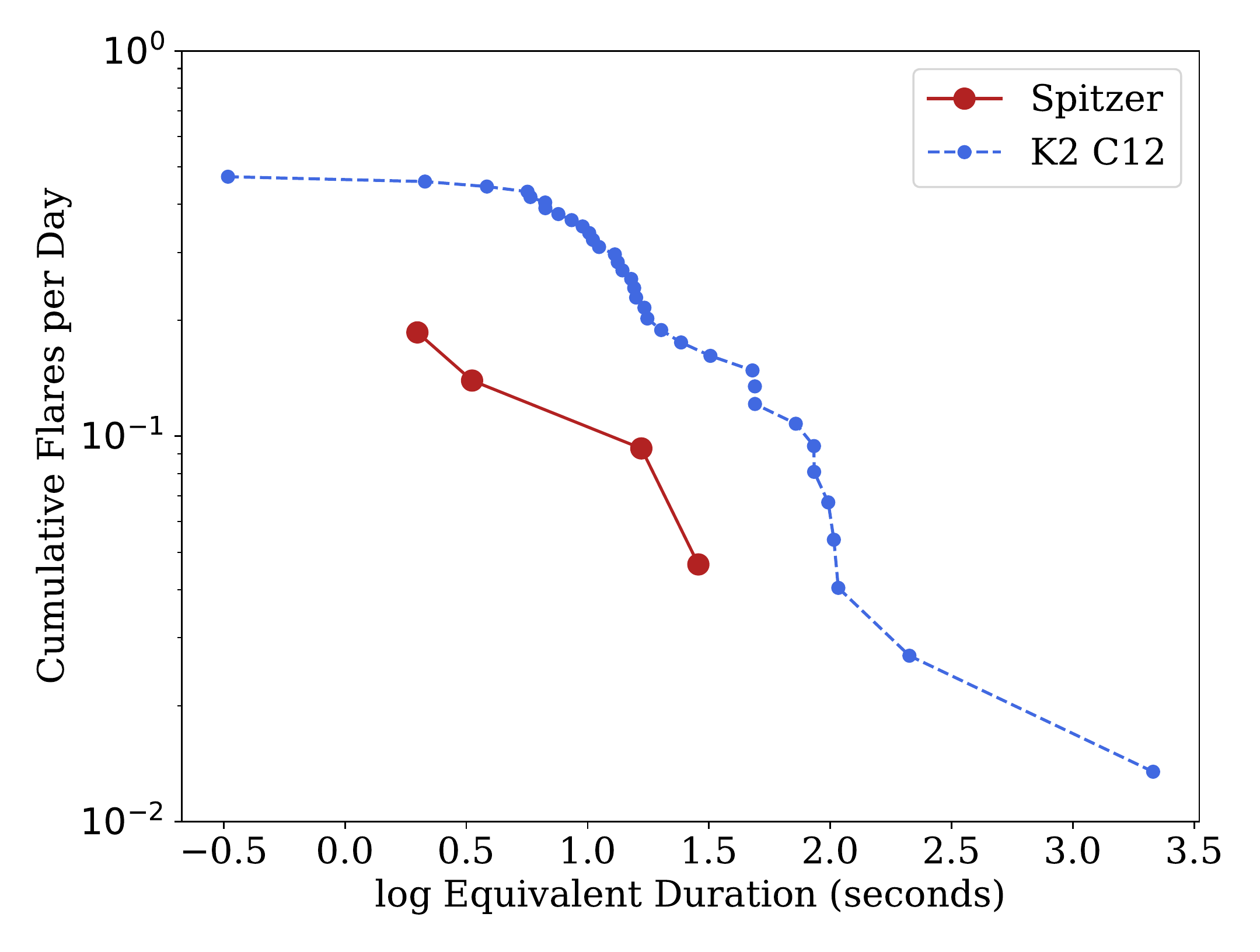}
\caption{Top: Example flares from Spitzer (left) and K2 (right) from the M8 dwarf TRAPPIST-1. 
Bottom: Cumulative flare frequency distributions for both Spitzer and K2 flare events in units of Equivalent Duration, which can be converted to event energies by multiplying by the stellar quiescent luminosity in each bandpass. 
\label{fig:1}}
\end{center}
\end{figure*}

Infrared stellar flares have been observed now, most notably in the Spitzer observations of TRAPPIST-1. This system is a nearby M8 star, hosting 7 transiting exoplanets discovered with ground- and space-based photometry \citep{gillon2016,gillon2017}. Four individual flare events are apparent in the Spitzer 4.5$\mu$m data\footnote{\url{https://github.com/jradavenport/trappist1_IRflares}} -- two more than reported by \citet{gillon2017}.
Follow-up data from \Kepler/K2 able revealed the orbital period of the outer planet \citep{luger2017}, and recovered many flares for this active star \citep{vida2017}. 
Interestingly, both the IR and the optical data contain flares that partially overlap transits: TRAPPIST-1b in the IR from \citet{gillon2017} and TRAPPIST-1h in the optical from \citet{luger2017}.
Figure \ref{fig:1} shows examples of isolated flares, as well as the power-law occurrence frequency distribution of flare events in both the optical and IR.

Since flares are common for low-mass stars, and can occur even for ``inactive'' stars \citep[e.g.][]{hawley2014}, I suggest that future surveys include optical monitoring during infrared transit observations to help calibrate out the effects of flares. Since flares are significantly higher amplitude in the optical \citep{davenport2012}, this monitoring does need not be at the same photometric precision as the IR, and can likely be done from the ground with modest aperture telescopes.

\acknowledgments

JRAD is supported by an NSF Astronomy and Astrophysics Postdoctoral Fellowship under award AST-1501418.


\end{document}